\documentstyle[12pt,a4]{article}

\begin{document}
\thispagestyle{empty}
\vspace*{-1.5cm}

\begin{center}
{\large A new look at the RST model}
\\[14mm]
{\large Jian-Ge Zhou\raisebox{0.8ex}{\small a},
F. Zimmerschied\raisebox{0.8ex}{\small a},
J.--Q. Liang\raisebox{0.8ex}{\small a,b} and \\
H.~J.~W. M\"uller--Kirsten\raisebox{0.8ex}{\small a} }
\\[1cm]
{\it \raisebox{0.8ex}{\small a} Department of Physics, \\
University of Kaiserslautern, P.\ O.\ Box 3049, D--67653 Kaiserslautern,
Germany \\[5mm]
\raisebox{0.8ex}{\small b} Institute of Theoretical Physics, \\
Shanxi University Taiyuan, Shanxi 030006, P.\ R.\ China \\
and \\
Institute of Physics, \\
Academia Sinica, Beijing 100080, P.\ R.\ China }
\\ \vfill
{\bf Abstract}
\end{center}
The RST model is augmented by the addition of a scalar field and
a boundary term so that it is well-posed and local. Expressing the RST
action in terms of the ADM formulation, the constraint
structure can be analysed completely. It is shown that from the view point
of local field theories, there exists a hidden dynamical field $\psi_1$ in
the RST model. Thanks to the presence of this hidden dynamical field, we
can reconstruct the closed algebra of the constraints which guarantee the
general invariance of the RST action. The resulting stress tensors T$_{\pm\pm}$
are recovered to be true tensor quantities. Especially, the part of the
stress tensors for the hidden dynamical field $\psi_1$ gives the precise
expression for $t_{\pm}$. At the quantum level, the cancellation condition for
the total central charge is reexamined. Finally, with the help of the
hidden dynamical field $\psi_1$, the fact that the semi-classical static
solution of the RST model has two independent parameters (P,M), whereas for
the classical CGHS model there is only one, can be explained.
\newpage
\indent
With the advent of the model proposed by Callan, Giddings, Harvey
and Strominger (CGHS) \cite{1}, dilaton gravity in two dimensions has been widely
recognized as an excellent arena in which a variety of fundamental issues
in quantum gravity can be discussed, especially those concerning quantum
properties of a black hole. Indeed now a large body of literature on the
CGHS model and its variants is available, and the notable model in the study
of the black hole evaporation problem is the Russo-Susskind-Thorlacious (RST)
model which admits physically sensible evaporating black hole solutions \cite{2}.
The RST model has been considered a theoretical laboratory for the study
of Hawking radiation \cite{2,3}, black hole entropy \cite{4,5,6}, critical phenomena
\cite{7,8} and so on.

However, until now, there are some problems which are still unclear in the RST
model. For example, the RST action is manifestly invariant under the diffeomorphism
transformation, so the constraints should form the closed algebra at the classical
level, i.e., ought to be first-class, which guarantees the general covariance
or invariance of the theory. Nevertheless, as is well known, the stress
tensors T$_{\pm\pm}$ in the earlier semiclassical approach do not transform as
tensors but rather as projective connections, which means that under a
conformal change of coordinates T$_{\pm\pm}$ pick up an extra term equal to
$-\kappa/2$ times the schwarzian derivative of the transition function at the
classical level \cite{9}. As a result, the Poisson brackets of the constraints
have the classical central extension, that is, the closed algebra of the
constraints is destroyed \cite{10}, which explicitly contradicts the fact that the
RST action is invariant with respect to diffeomorphism transformations.

Usually, arbitrary functions (more precisely, projective connections)
t$_{\pm}$ are added to T$_{\pm\pm}^{\rho,\phi}$ by hand, and under a conformal
change of coordinates t$_{\pm}$ are assumed to pick up $\kappa/2$ times the
schwarzian derivative, so that T$_{\pm\pm}$ + t$_{\pm}$ are true tensor
quantities. Since t$_{\pm}$ are introduced by hand, their precise meanings
are implicit, so t$_{\pm}$ have various physical explanations. For
instance, in \cite{11} t$_{\pm}$ are explained as the stress tensors for the ghost
sector, and their central charges are equal to 26, whereas in \cite{3,9} t$_{\pm}$
are considered as the result of the nonlocality of the Polyakov term, and
the corresponding central charges are 12$\kappa$. So, until now, it is not
clear how these conflicts could be reconciled in a consistent way.

In the present paper, the RST model is first dicussed from the viewpoint of
the Dirac quantization method so as to solve the above mentioned problems.
Since there is a nonlocal term in the RST model, the RST Lagrangian must
first be localized so that Dirac quantization can be performed. For this
purpose, the scalar field $\chi$ and the boundary term are introduced in order
that the reformulated RST model is well-posed and local \cite{6,15}. Expressing the
RST action in terms of the ADM formulation \cite{12,13,14},
the constraint structure can be easily analysed. It is found that there are
four first-class constraints in the RST model, and two of these generate the
well-known Virasoro algebra without classical central charge. At the quantum
level, the cancellation condition for the total central charge is reexamined.
Three types of measures are discussed and the corresponding results are
obtained. By the Hamiltonian constraint analysis, it is shown that except for
the N scalar matter degrees of freedom, the true physical degrees of freedom
for gravity, the dilaton and the new field $\chi$, are nonzero. From the
viewpoint of the local field theories, there is a hidden dynamical field in
the RST model, which was omitted in the usual semiclassical approach.
Exploiting the equations of motion, the stress tensors T$_{\pm\pm}$ can be
derived from our original constraints ${\cal H}_{\pm}$. In comparison with the
known results, it is found that just the stress tensors of the hidden dynamical
field $\psi_1$ give the precise expression for t$_{\pm}$. Thus we conclude
that the previous semiclassical approach is intrinsically inconsistent due
to the omission of this hidden dynamical field $\psi_1$, which results in the
above mentioned conflicts. Finally, with the help of the hidden dynamical
field $\psi_1$, the fact that the semi-classical static solution for the RST
model has two
independent parameters (P,M), whereas for the classical CGHS model it has only
one, can be well elucidated.

\indent

We now consider the RST model with the action \cite{2}

\begin{equation}
S = \frac{1}{2\pi}\int_{\cal M} d^2x\sqrt{-g}\left[e^{-2\phi}(R+4(\nabla\phi)^2
+4\lambda^2) - \frac12 \sum^N_{i=1} (\nabla f_i)^2 - \frac{\kappa}{4} (R
\frac{1}{\nabla^2}R + 2\phi R)\right]
\label{1}
\end{equation}
where g$_{\mu\nu}$ is the metric on the 2D manifold $\cal M$, R is its curvature
scalar, $\phi$ is the dilaton field, and the $f^i$, i = 1,...,N, are N scalar
matter fields. The nonlocal term $R\frac{1}{\nabla^2}R$ comes from the
familiar conformal anomaly. The local and covariant term $2\phi R$ is to
preserve the simple form of the current $j^{\mu} = \partial^{\mu}(\phi-\rho)$, with
$\partial_{\mu}j^{\mu} = 0$. The coefficient $\kappa$ has to be positive, since
in the case of $\kappa$ being negative, there is no singularity in
gravitational collapse \cite{2}. Obviously, Eq. (\ref{1}) is invariant with respect
to the diffeomorphism transformation.

According to Ref. \cite{15}, one can introduce an independent scalar
field $\chi$
to localize the conformal anomaly term, and add a boundary term to define
the variational problem properly. Then Eq. (\ref{1}) becomes \cite{15}

\begin{eqnarray}
S& = & \frac{1}{2\pi}\int d^2x\sqrt{-g}\{R\tilde{\chi}+4 [(\nabla\phi)^2
+\lambda^2]e^{-2\phi} - \frac{\kappa}{4} g^{\alpha\beta}\partial_{\alpha}\chi
\partial_{\beta}\chi \nonumber \\
& & - \frac12 \sum^{N}_{i=1} (\nabla f_i)^2 \} - \frac{1}{\pi} \int d\Sigma
\sqrt{-h} K \tilde{\chi}
\label{2}
\end{eqnarray}
where $\tilde{\chi} = e^{-2\phi} - \frac{\kappa}{2}(\phi-\chi)$, h is
the induced metric on the boundary of $\cal M$ (assumed spacelike), and K
is the mean extrinsic curvature of $\partial \cal M$. As in (3+1)-dimensional
gravity, the boundary term serves to eliminate second time derivatives of the
metric from the action which are contained in R.

Following the ADM formulation, the metric can be parametrized as follows 
\cite{12,13,14}:

\begin{equation}
g_{\mu\nu} = e^{2\rho} \hat{g}_{\mu\nu}
\label{3a}
\end{equation}

\begin{equation}
\hat{g}_{\mu\nu} = \left(\begin{array}{cc}
-\sigma^2+\theta^2 & \theta\\
\theta & 1 \end{array}\right)
\label{3b}
\end{equation}
where $\sigma(x)$ and $\theta(x)$ are lapse and shift functions respectively,
and we factor out the conformal factor e$^{2\rho}$.

In terms of this parametrization, the action (\ref{2}) can be written as

\begin{eqnarray}
S & = & \frac12 \int d^2x \sqrt{\hat{g}} \Big\{\hat{R}\tilde{\chi} +
2\hat{g}^{\alpha\beta} \partial_\alpha \tilde{\chi} \partial_\beta\rho -
2\hat{g}^{\alpha\beta} \partial_\alpha \phi \partial_\beta e^{-2\phi}
\nonumber \\
& & + 4\lambda^2 e^{2(\rho-\phi)} - \frac{\kappa}{4} \hat{g}^{\alpha\beta}
\partial_\alpha \chi \partial_\beta\chi - \frac12 \sum^{N}_{i=1} \hat{g}
^{\alpha\beta} \partial_\alpha f_i \partial_\beta f_i \Big\} \nonumber \\
& & - \int d \Sigma \sqrt{\pm h} K\tilde{\chi}
\label{4}
\end{eqnarray}
where $\hat{R}$ is the curvature scalar for $\hat{g}_{\mu\nu}$, and for
simplicity, the factor $\pi^{-1}$ in front of action (\ref{2}) has been omitted.

If we introduce momenta $\pi_{\rho}$, $\pi_{\phi}$, $\pi_{\chi}$ respectively
for the fields $\rho, \phi, \chi$, the Hamiltonian would become so complicated
that we cannot quantize the theory. Thus we need a field redefinition to
diagonalize the kinetic term of action (\ref{4}), which is first given by

\begin{eqnarray}
\psi_0 & = & \frac{1}{\sqrt{\kappa}} e^{-2\phi} - \frac{\sqrt{\kappa}}{2}\phi
+ \sqrt{\kappa}\rho \nonumber \\
\psi_1 & = & -\frac{\sqrt{\kappa}}{2}\chi
+ \sqrt{\kappa}\rho \nonumber \\
\psi_2 & = & \frac{1}{\sqrt{\kappa}} e^{-2\phi} +\frac{\sqrt{\kappa}}{2}\phi
\label{5}
\end{eqnarray}
Here we should point out that the physical value of $\psi_2$ is restricted,
i.e., it is a non-negative quantity. If this restriction is ignored, the semi-
classical solution of the model is unstable \cite{16}. The black holes
radiate forever at a fixed rate and the Bondi mass tends to negative infinity.
This feature will not be changed by the addition of the auxilliary field to the
model \cite{17}. Then we have

\begin{eqnarray}
S & = & \int d^2x \Big\{ \frac{\sqrt{\kappa}}{\sigma} (\dot{\psi}_0
 - \dot{\psi}_1)\theta^{\prime} + \frac{\sqrt{\kappa}}{\sigma} (\psi^{\prime}_0
- \psi^{\prime}_1)(\sigma\sigma^{\prime}-\theta\theta^{\prime}) \nonumber \\
& & +\frac12 \sigma \hat{g}^{\alpha\beta} \partial_{\alpha} \psi_{\mu}
\partial_{\beta} \psi_{\nu} \eta^{\mu\nu} + 2\lambda^2\sigma
e^{\frac{2}{\sqrt{\kappa}} (\psi_0-\psi_2)} - \frac14 \sigma \sum^{N}_{i=1}
\hat{g}^{\alpha\beta} \partial_\alpha f_i \partial_\beta f_i \Big \}
\label{6}
\end{eqnarray}
where $\mu,\nu = 0,1,2$ with $\eta^{\mu\nu} = (1,-1,-1)$, $\hat{g}^{00} = -\sigma^2$,
$\hat{g}^{01} = \hat{g}^{10} = \theta\sigma^{-2}$, $\hat{g}^{11} = (\sigma^2
- \theta^2)\sigma^{-2}$. In the above, dots and primes denote differentiation
with respect to time and space respectively. The canonical momenta associated
with the fields
$\{\sigma,\theta,\psi_{\mu}, f_i \}$ are

\begin{eqnarray}
P_{\sigma} & = & 0  \label{7a} \\
P_{\theta} & = & 0
\label{7b} \\
P_0 & = & - \frac {\dot{\psi}_0}{\sigma} + \frac {\theta\psi^{\prime}_0}{\sigma}
+ \frac {\sqrt{\kappa}\theta^{\prime}}{\sigma} \label{7c} \\
P_1 & = & \frac {\dot{\psi}_1}{\sigma} -\frac {\theta\psi^{\prime}_1}{\sigma}
- \frac {\sqrt{\kappa}\theta^{\prime}}{\sigma} \label{7d} \\
P_2 & = & \frac {\dot{\psi}_2}{\sigma} - \frac {\theta\psi^{\prime}_2}{\sigma}
\label{7e}\\
\pi_i & = & \frac{\dot{f}_i}{2\sigma} - \frac{\theta f^{\prime}_i}{2\sigma}
\label{7f}
\end{eqnarray}

with

\begin{eqnarray}
\{\sigma(x),P_{\sigma}(y)\} & = & \{\theta(x),P_{\theta}(y)\} = \delta(x-y)
\nonumber \\
\{\psi_{\mu}(x),P_{\nu}(y)\} & = & \delta_{\mu\nu} \delta(x-y) \nonumber \\
\{f_i(x),\pi_j(y)\} & = & \delta_{ij} \delta(x-y)
\label{8}
\end{eqnarray}
Clearly (\ref{7a}) and (\ref{7b}) are primary constraints and $\sigma(x)$ and $\theta(x)$
play the role of Lagrange multipliers. The canonical Hamiltonian, up to
surface terms, is

\begin{equation}
{\cal H}_c = \int dx (\sigma {\cal H}_{\sigma} + \theta {\cal H}_{\theta})
\label{9}
\end{equation}
where

\begin{eqnarray}
{\cal H}_{\sigma} & = & - \frac12 (P^2_0 + \psi^{\prime2}_{0}) + \frac12
(P^2_1 + \psi^{\prime2}_{1}) + \frac12 (P^2_2 + \psi^{\prime2}_{2})
\nonumber \\
& & + \sqrt{\kappa} (\psi^{\prime\prime}_0 - \psi^{\prime\prime}_1) -
2\lambda^2 e^{\frac{2}{\sqrt{\kappa}} (\psi_0 - \psi_2)}+ \sum^{N}_{i=1}
(\pi^2_i + \frac14 f^{\prime2}_i) = 0
\label{10a}
\end{eqnarray}

\begin{equation}
{\cal H}_{\theta} = P_0\psi^{\prime}_0 + P_1\psi^{\prime}_1 + P_2\psi^{\prime}_2
- \sqrt{\kappa} (P^{\prime}_0 + P^{\prime}_1) + \sum^{N}_{i=1} \pi_i f^{\prime}
_i = 0
\label{10b}
\end{equation}
are secondary constraints. $\cal H_{\theta}$ is the generator of spatial
diffeomorphisms, but $\cal H_{\sigma}$ does not exactly correspond to the
generator of temporal diffeomorphisms \cite{18}. Since the constraint ${\cal H}_{\sigma}$
is nonlinear in the momenta, it does not generate a transformation which
corresponds to a symmetry of the corresponding Lagrangian system. Rather
it is responsible for the dynamics of the system. On the other hand, the
transformation generated by $\cal H_{\sigma}$ is indeed a symmetry of the
Hamiltonian system (which cannot be identified with a Lagrangian symmetry
in gravity theory) \cite{18}. However, all Lagrangian symmetries can be recovered
in the Hamiltonian formalism only if we consider the transformation generated
by $\cal H_{\sigma}$ in a very special combination with a particular ``trivial"
transformation \cite{18}.

We now calculate the Poisson brackets of the constraints $\cal H_{\sigma}$, $\cal H_{\theta}$,
and after a series of steps, we have

\begin{equation}
\{{\cal H}_{\sigma}(x),{\cal H}_{\sigma}(y)\} = [{\cal H}_{\theta}(x) +
{\cal H}_{\theta}(y)] \partial^x_1\delta (x^1 - y^1)
\label{11a}
\end{equation}

\begin{equation}
\{{\cal H}_{\theta}(x),{\cal H}_{\theta}(y)\} = [{\cal H}_{\theta}(x) +
{\cal H}_{\theta}(y)] \partial^x_1\delta (x^1 -
y^1)
\label{11b}
\end{equation}

\begin{equation}
\{{\cal H}_{\sigma}(x),{\cal H}_{\theta}(y)\} = [{\cal H}_{\sigma}(x) +
{\cal H}_{\sigma}(y)] \partial^x_1\delta (x^1 - y^1)
\label{11c}
\end{equation}

Eqs. (\ref{11a}-\ref{11c}) show that $\cal H_{\sigma}, \cal H_{\theta}$ form a
closed algebra under Poisson brackets, that is, they are first-class
constraints at the classical level. Here we emphasize that thanks to the
existence of the scalar field $\chi$, the closed algebra is recoverd,
which guarantees the general invariance of the RST action.

It is obvious that the total number of degrees of freedom is $5 + N (i.e., \sigma,
\theta, \psi_{\mu}, f_i)$, while there are four first-class constraints,
so the true number of physical degrees of freedom is 1+N, i.e., $(5+N) - (2+2)
= 1+N$. From the view point of local field theories, we find that except for
the N scalar matter fields, there is another dynamical field $\psi_1$, and
we call it a hidden dynamical field, which was omitted in the previous
semiclassical approach. In the present case, due to the presence of this
hidden dynamical variable, the constraints $\cal H_{\sigma}, \cal H_{\theta}$
are recovered to be first-class.

According to Dirac's algorithm, the conditions of a physical state $\Psi$ can
be expressed as

\begin{eqnarray}
{\cal H}_{\sigma}\Psi & = & \Big\{ - \frac12 (P^2_0 + \psi^{\prime2}_0) +
\frac12 (P^2_1 + \psi^{\prime2}_1)+ \frac12 (P^2_2 + \psi^{\prime2}_2) +
\sqrt{\kappa}(\psi''_0 - \psi''_1)
\nonumber \\
& & -2\lambda^2 e^{ \frac{2}{\sqrt{\kappa}} (\psi_0-\psi_2)} + \sum^{N}_{i=1}
(\pi^2_i + \frac14 f^{\prime2}_i) \Big\} \Psi = 0
\label{12a}
\end{eqnarray}

\begin{equation}
{\cal H}_{\theta}\Psi = \Big\{ P_0\psi^{\prime}_0 + P_1\psi^{\prime}_1 +
P_2\psi^{\prime}_2 - \sqrt{\kappa} (P_0^{\prime}+P^{\prime}_1) +
\sum^{N}_{i=1} \pi_i f^{\prime}_i \Big\} \Psi = 0
\label{12b}
\end{equation}

Eqs. (\ref{12a},\ref{12b}) are just modified versions of the Wheeler-DeWitt equation [19,20].
Here we note that the constraints $P_{\sigma}=0$ and $P_{\theta}=0$ require the
wave functional $\Psi$ to be independent of the Lagrange multipliers $\sigma(x)$
and $\theta(x)$.
So the physical state will have the form

\begin{equation}
\Psi = \Psi(\psi_\mu,f_i)
\label{13}
\end{equation}
in the functional Schr\"odinger representation.

Owing to the algebra (\ref{11a}-\ref{11c}) being isomorphic to two commuting copies of the
1D diffeomorphism algebra, we can construct the constraints in terms of the
light cone ones:

\begin{eqnarray}
{\cal H}_{\pm} & = & \frac12 ({\cal H}_{\sigma} \pm {\cal H}_{\theta})
= - \frac14 (P_0 \mp \psi^{\prime}_0)^2 + \frac14 (P_1 \pm \psi^{\prime}_1)^2
+ \frac14 (P_2 \pm \psi^{\prime}_2)^2 \nonumber \\
& &+\frac{\sqrt{\kappa}}{2} (\mp P^{\prime}_0 + \psi^{\prime\prime}_0) -
\frac{\sqrt{\kappa}}{2} (\pm P^{\prime}_1 + \psi^{\prime\prime}_1) - \lambda^2
e^{\frac{2}{\sqrt{\kappa}} (\psi_0-\psi_2)} \nonumber \\
& & + \frac12 \sum^{N}_{i=1} (\pi_i \pm \frac12 f^{\prime}_i)^2 = 0
\label{14}
\end{eqnarray}
>From Eqs. (\ref{11a}-\ref{11c},\ref{14}), we immediately recognize $\cal H_{\pm}$ obeying the
Virasoro algebra \cite{21}.

In the conformal gauge (which means $\sigma = 1, \theta = 0)$, $g_{++} = g_{--}
= 0$, $g_ {+-} = - \frac12 e^{2\rho}$, the action (\ref{6}) can be written as

\begin{eqnarray}
S & = & \int d^2x \Big[ - \partial_+\psi_0 \partial_-\psi_0 +
\partial_+\psi_1\partial_-\psi_1 + \partial_+\psi_2\partial_-\psi_2 +
\lambda^2 e^{\frac{2}{\sqrt{\kappa}} (\psi_0-\psi_2)} \nonumber \\
& & + \frac12 \sum^{N}_{i=1} \partial_+f_i\partial_-f_i \Big]
\label{15}
\end{eqnarray}
and the constraints (\ref{14}) become

\begin{eqnarray}
{\cal H}_{\pm} & = & - \partial_{\pm}\psi_0 \partial_{\pm}\psi_0 + \sqrt{\kappa}
\partial^2_{\pm} \psi_0 +\partial_{\pm}\psi_2 \partial_{\pm}\psi_2
\nonumber \\
& & + \frac12 \sum^{N}_{i=1} \partial_{\pm}f_i \partial_{\pm}f_i +
\partial_{\pm}\psi_1 \partial_{\pm}\psi_1 - \sqrt{\kappa}
\partial^2_{\pm}\psi_1 \nonumber \\
& & - \sqrt{\kappa}\partial_+\partial_-\psi_0 +
\sqrt{\kappa}\partial_+\partial_-\psi_1 - \lambda^2
e^{\frac{2}{\sqrt{\kappa}} (\psi_0-\psi_2)} = 0
\label{16}
\end{eqnarray}

The equations of motion derived from action (\ref{15}) are

\begin{eqnarray}
\partial_+\partial_-\psi_0 & = & - \frac{\lambda^2}{\sqrt{\kappa}}
e^{\frac{2}{\sqrt{\kappa}} (\psi_0-\psi_2)} \label{17a} \\
\partial_+\partial_-\psi_1 & = & 0 \label{17b} \\
\partial_+\partial_-\psi_2 & = & - \frac{\lambda^2}{\sqrt{\kappa}}
e^{\frac{2}{\sqrt{\kappa}} (\psi_0-\psi_2)} \label{17c} \\
\partial_+\partial_-f_i & = & 0
\label{17d}
\end{eqnarray}
With Eqs. (\ref{17a}) and (\ref{17b}), the constraints $\cal H_{\pm}$ can be reduced to

\begin{eqnarray}
{\cal H}_{\pm} & = & - \partial_{\pm}\psi_0 \partial_{\pm}\psi_0 + \sqrt{\kappa}
\partial^2_{\pm}\psi_0 + \partial_{\pm}\psi_2 \partial_{\pm}\psi_2 \nonumber \\
& & + \frac12 \sum^N_{i=1} \partial{\pm}f_i\partial_{\pm}f_i +
\partial_{\pm}\psi_1 \partial_{\pm}\psi_1 - \sqrt{\kappa}
\partial^2_{\pm}\psi_1 = 0
\label{18}
\end{eqnarray}
In comparison with previous results \cite{2,3,9,11}, $\cal H_{\pm}$
are nothing but the stress tensors
T$_{\pm\pm}$ with added contributions

\begin{equation}
t_{\pm} = \partial_{\pm}\psi_1 \partial_{\pm}\psi_1 - \sqrt{\kappa}
\partial^2_{\pm}\psi_1
\label{19}
\end{equation}

Eqs. (\ref{18},\ref{19}) show that T$_{\pm\pm}$ are true tensor, and under a
conformal change of coordinates t$_{\pm}$ indeed pick up $- \kappa /2$ times
the schwarzian derivative. In our derivation, t$_{\pm}$ appear in a natural way,
as a matter of fact, t$_{\pm}$ are just the stress tensors for the hidden
dynamical field $\psi_1$. If the hidden dynamical field $\psi_1$ is omitted,
the above mentioned conflicts will arise, that is, the original stress
tensors T$_{\pm\pm}$ will turn out to be nontensor, and the algebra (\ref{11a}-\ref{11c}) will
not be closed.

\indent

>From the above discussion, we find that due to the presence of the hidden
dynamical field $\psi_1$, the constraints form the closed algebra without
classical central extension. At the quantum level, we now apply the Bilal--
Callan method \cite{9} to analyse the quantum central charge. From the expression
for the stress tensors T$_{\pm\pm}$

\begin{eqnarray}
T_{\pm\pm} & = & - \partial_{\pm}\psi_0\partial_{\pm}\psi_0 + \sqrt{\kappa}
\partial^2_{\pm}\psi_0 + \partial_{\pm}\psi_2\partial_{\pm}\psi_2 \nonumber \\
& & +\frac12 \sum^{}_{i=1} \partial_{\pm}f_i\partial_{\pm}f_i +
\partial_{\pm}\psi_1\partial_{\pm}\psi_1 - \sqrt{\kappa}
\partial^2_{\pm}\psi_1 = 0
\label{20}
\end{eqnarray}
one can easily obtain the cancellation condition for the total quantum
central charge:

\begin{eqnarray}
C & = & C_{\psi_0} + C_{\psi_1} + C_{\psi_2} + C_M + C_{ghost} \nonumber \\
& = & (1-12\kappa) + (1+12\kappa) + 1 + N - 26 = 0
\label{21}
\end{eqnarray}
with

\begin{equation}
N = 23
\label{22}
\end{equation}
At first sight, this result seems somewhat surprising, Eq. (\ref{21}) cannot
determine the value of $\kappa$, but gives the restriction on N. This
is because the stress tensors in the present case have no classical
central charge, which is similar to the classical CGHS model where the
stress tensors are true tensor, so the condition without conformal anomaly
in the CGHS model is N = 24 \cite{16}. Our result can also be understood from
the measure definition. Suppose we start with action (\ref{6}) and take $\psi_{\mu},f_i$
as our fundamental fields. The condition (\ref{21}) then means the functional
measures are defined by the following norms:

\begin{eqnarray}
\parallel\delta\psi_\mu\parallel^2_{\hat{g}} & = & \int d^2x \sqrt{-\hat{g}}
\delta^{\mu\nu} \delta\psi_{\mu}\delta\psi_{\nu} \\
\parallel\delta f_i\parallel^2_{\hat{g}} & = & \int d^2x \sqrt{-\hat{g}}
\delta^{ij} \delta f_i\delta f_j
\label{23}
\end{eqnarray}

If we take the functional measures for the fields $g_{\mu\nu},\phi,f_i$ to be
those defined by the norms

\begin{equation}
\parallel\delta g\parallel^2_g = \int d^2x \sqrt{-g} g^{\alpha\gamma}
g^{\beta\delta} (\delta g_{\alpha\beta} \delta g_{\gamma\delta} +
\delta g_{\alpha\gamma}\delta g_{\beta\delta})
\label{24a}
\end{equation}

\begin{equation}
\parallel\delta\phi\parallel^2_g = \int d^2x \sqrt{-g} (\delta\phi)^2
\label{24b}
\end{equation}

\begin{equation}
\parallel\delta f_i\parallel^2_g = \int d^2x \sqrt{-g} \delta^{ij}
\delta f_i \delta f_j
\label{24c}
\end{equation}
and consider the classical CGHS action as our starting point, one might 
argue \`a
la David, Distler and Kawai (DDK) \cite{22} about the measure in the path
integral; then the condition for a vanishing central charge is

\begin{equation}
\kappa = \frac{N-24}{12}
\label{25}
\end{equation}
which is just the approach adopted in Refs. \cite{11,23}. However, if we replace
(\ref{24b}) by \cite{24}

\begin{equation}
\parallel\delta\eta\parallel^2_g = \int d^2x \sqrt{-g} (\delta\eta)^2
\label{26}
\end{equation}
with
\begin{equation}
\eta = e^{- \phi}
\label{27}
\end{equation}
i.e., consider $e^{-\phi}$ as original field, then the corresponding
condition becomes \cite{10,24}

\begin{equation}
\kappa = \frac{N - \frac{51}{2}}{12}
\label{28}
\end{equation}

Generally speaking, different types of measures used will result in different
conditions for the total quantum central charge to vanish \cite{24}.

The general $f_i = 0$ solution for the classical CGHS model is \cite{1}

\begin{equation}
e^{-2\phi} = e^{-2\rho} = \frac{M}{\lambda} - \lambda^2x^+x^-
\label{29}
\end{equation}
where only one global parameter M exists. The semi-classical static
solution of the RST model is \cite{2}

\begin{equation}
\frac{\sqrt{\kappa}}{2} \phi + \frac{e^{-2\phi}}{\sqrt{\kappa}} =
- \frac{\lambda^2x^+x^-}{\sqrt{\kappa}} + P \sqrt{\kappa}
\ln (-\lambda^2x^+x^-) + \frac{M}{\lambda\sqrt{\kappa}}
\label{30}
\end{equation}
where P and M parametrize different solutions, i.e., there are two
global and independent parameters P,M. However, as we know, the equations
of motion for both models are differential equations of the same order.
One may wonder why both models do not have the same number of global parameters.
The reason is that the classical CGHS model has no local degrees of freedom
\cite{15} when
$f_i$ are zero, whereas the RST model has a hidden dynamical field $\psi_1$
which is responsible for the parameter P.

>From Eqs. (\ref{17a}-\ref{17d},\ref{18}), we have \cite{2,11}

\begin{eqnarray}
\psi_0 = \psi_2 & = & - \frac{\lambda^2x^+x^-}{\sqrt{\kappa}} +
\Big[ \int dx^+ \int dx^+ \Big( \partial^2_+\psi_1 - \frac{1}{\sqrt{\kappa}}
(\partial_+\psi_1)^2 \Big) \nonumber \\
& & + \int dx^- \int dx^- \Big( \partial^2_-\psi_1 - \frac{1}{\sqrt{\kappa}}
(\partial_-\psi_1)^2 \Big) \Big] + \frac{m}{\lambda \sqrt{\kappa}}
\label{31}
\end{eqnarray}
Eq. (\ref{17b}) shows that $\psi_1$ satisfies a free massless scalar field
equation with solution $\psi_1 = \psi^+_1(x^+) + \psi^-_1(x^-)$, so we have
the freedom to choose

\begin{equation}
\psi^+_-(x^+) = c \ln(\lambda x^+), \psi^-_1(x^-) = c \ln(-\lambda x^-)
\label{32}
\end{equation}
where c is an arbitrary constant.
Then Eq. (\ref{31}) reduces to

\begin{equation}
\psi_0 = \psi_2 = - \frac{\lambda^2x^+x^-}{\sqrt{\kappa}} + P\sqrt{\kappa}
\ln (-\lambda^2x^+x^-) + \frac{M}{\lambda\sqrt{\kappa}}
\label{33}
\end{equation}
with
\begin{equation}
P = \left( c + \frac{c^2}{\sqrt{\kappa}} \right)/\sqrt{\kappa}
\label{34a}
\end{equation}

\begin{equation}
M = m - P\sqrt{\kappa} \ln \lambda^2
\label{34b}
\end{equation}
Eqs. (\ref{33},\ref{34a}) show that the hidden dynamical field $\psi_1$
induces the parameter P. This result is consistent with the fact that in
the semi-classical CGHS model including the conformal anomaly, the static
solution (which can be studied numerically) have two parameters, one of
which corresponds to the energy density in the asymptotic region.

\indent

In summary, we have reconstructed the closed algebra for the constraints with
the help a of hidden dynamical field $\psi_1$, and the resulting stress
tensors T$_{\pm\pm}$ are true tensor. If the hidden dynamical field $\psi_1$ is omitted
as in the usual semiclassical approach, the theory will be inconsistent.
For example, under a conformal change of coordinates the stress tensors
T$_{\pm\pm}$ will pick up an extra term equal to $-\kappa/2$ times the
schwarzian derivative of the transition function at the classical level.
Thus the Poisson brackets of the constraints will have the classical extension,
i.e., the closed algebra of the constraints (\ref{11a}-\ref{11c}) will be destroyed,
in contradiction with the fact that the RST action is manifestly invariant
under the diffeormorphism transformation. Thanks to the existence of the
hidden dynamical field, the stress tensors t$_{\pm}$ can be endowed with
precise meaning, and the contradictions mentioned in the introduction can be
resolved in a perfect manner. Now with the diagonalized action (\ref{6})
and a clear constraint structure at hand, we hope to understand some quantum
physics in the strong coupling regime with the path integral approach \cite{25},
and meanwhile we can also shed some new light on the physical meaning of the
hidden dynamical field \cite{26}. Another aspect of interest is to solve the
Wheeler-DeWitt equation (\ref{12a},\ref{12b}) in the functional Schr\"odinger
representation to obtain the physical wave functional $\Psi$, from which we
can obtain the entropy of the RST model \cite{27} in order to understand the
origin of the black hole entropy more deeply. These problems are presently under
investigation and we hope to be able to report our progress elsewhere.

\section*{Acknowledgement}
This work was supported in part by the European Union under the Human Capital
and Mobility programme. One of us (J.-G. Z.) thanks the Alexander von Humboldt
Foundation for financial support in the form of a research fellowship.

\end{document}